\input aa.cmm
\voffset=-1 truecm
\input psfig.tex
\overfullrule=0 pt
\MAINTITLE{Structural classification of galaxies in clusters}
\AUTHOR{S. Andreon@1@,@2\FOOTNOTE{Present adress: Observatoiro di Capodimonte, via Moiariello 16, 80131 Naples, Italy, e-mail: andreon at cerere.na.astro.it}
and E. Davoust@2}
\INSTITUTE{@1 Istituto di Fisica Cosmica del CNR, via
            Bassini 15, 20133 Milano, Italy\newline
           @2 CNRS-UMR 5572, Observatoire Midi-Pyr\'en\'ees, 14, Av. E.
            Belin, 31400 Toulouse, France, e-mail: davoust at obs-mip.fr\newline
                     }
\DATE{ Received May, 24, 1996; accepted September, 4, 1996} 

\ABSTRACT{
The traditional method of morphological classification, by visual
inspection of images of uniform quality and by reference to standards for
each type, is critically examined.  The rate of agreement among
traditional morphologists on the morphological type of galaxies is
estimated from published classification works, and is estimated at about
20 \%, when galaxies are classified into three bins (E, S0, S+Irr). 

The advantages of the quantitative method of structural classification 
for classifying galaxies in clusters are outlined.  This method is based on 
the isophotal analysis of galaxy images, and on the examination of 
quantitative structural parameters derived from this analysis, such as the 
profiles of luminosity, ellipticity and deviations from ellipticity of the 
galaxy.  

The structural and traditional methods are compared on a complete sample of 
190 galaxies in the Coma cluster.  The morphological types derived by both 
methods agree to within 15 or 20 \%, the same rate as among traditional 
morphologists alone, thus showing that our morphological classes do
coincide with the traditional ones.  The galaxies with discrepant types are 
mostly faint (mag$_B > 16.0$), or have features typical of spirals, but which 
have not been detected, noticed or taken into account by traditional 
morphologists.  The rate of agreement is also good for galaxies in a distant 
cluster ($z\sim 0.4$). 

The structural method, which requires an image quality adapted to the
difficulty of classifying a given galaxy, gives highly reproducible
results, never reached by traditional estimations of the morphological
type. Thus, the morphological types obtained with this method should be
preferred, even if their determination is more time and telescope
consuming, because they are less subjective, therefore more reproducible,
and based on images of adequate resolution.  The advantages of the method
are further demonstrated by new results on the properties of galaxies in
clusters.}

\THESAURUS{03(11.03.4 Coma cluster;  11.05.1; 11.06.2; 11.19.2)} 
\maketitle 
\titlea {Introduction}

Classification is the first task to be undertaken when exploring a new field. 
A good classification system should separate the bewildering diversity of 
observed shapes into a finite number of bins containing objects with specific 
physical properties, and thus provide a better understanding of the physical 
nature of the objects under investigation.  In order to do so, this system 
should be based on structural properties, and should ignore others, even if 
they are aesthetically pleasing.  The classification criteria should also
provide a non ambiguous assignment to a class for each object; more than one 
criterion per class may lead to two equally possible classifications for a 
given object. 

The morphological classification of galaxies first proposed by Hubble (1936)
has been universally adopted with little change and is still being used, 
because it does break galaxies into classes with specific physical properties.
Its application to galaxies in clusters has revealed a morphological 
segregation of galaxies which is probably a key element for understanding the 
formation and evolution of galaxies, and the investigation of the luminosity 
function for each morphological type should shed further light on this 
question. The refurbishing of the Hubble Space Telescope ({\it HST}) has 
renewed interest in the morphological classification of galaxies, as it will 
enable us to compare the morphological composition of nearby and distant (z 
$\simeq 0.2 - 0.4$) samples, and thus to obtain information on the evolution of 
galaxies. 

While the work of classifying galaxies has traditionally been done by visual 
inspection of images of galaxies, recent progress in the fields of digital 
detectors and image treatment by computer has given rise to new methods of 
investigation of galaxy images, such as isophotal analysis (e.g. Poulain, 
Nieto \& Davoust 1992, Michard \& Marchal 1993), which in turn has lead to 
further refinements of the classification system.  Such refinements might not 
necessarily bring out new physical idiosyncrasies of galaxies; indeed, it 
still remains to be shown that the dichotomy of elliptical galaxies into boxy 
and disky breaks these galaxies into two subclasses with physically distinct 
internal properties (Andreon 1996). The new perspectives opened by computer 
treatment of images should nevertheless be pursued, as computers eliminate 
part of the subjectivity in the task of classifying galaxies. 

In this paper, after analyzing the traditional morphological
classification system (Sect. 2) and its problems (Sect. 3 and 4), we
present a quantitative method for classifying galaxies in clusters, based
on the analysis of quantitative structural parameters, such as the
luminosity, ellipticity and $e_4$ profiles (Sect. 5).  We show that this
method is identical in spirit to the traditional system by comparing the
morphological types obtained by both methods for a large sample of
galaxies in the Coma cluster (Sect. 7) and for galaxies at high redshift
(Sect. 8).  The justification for preferring our method is that it has the
advantage of giving highly reproducible results and requires a uniform
resolution in the restframe of the galaxies, at least for the most
difficult cases, and that it has already demonstrated its merits (Sect.
6). 

\titlea{Traditional morphologists' classification}

\noindent
Most of our present knowledge of galaxy morphology is based on the pioneering 
works of a few observers who classified thousands of galaxies. The 
observational material available for the classification (better material 
allows a more detailed scheme) and the aims of the classification work 
generally govern the choice of the classification scheme. The details of these 
schemes depend on the authors; nevertheless, some points are common to most 
authors. 

\medskip

\item {1.} The morphological `system is usually defined by a set of standards
or prototypes' (Buta 1990), and the galaxies are classified according the
resemblance to these standards (although it is not always the case; see e.g.
Kormendy 1979, Kennicutt 1981 or Schombert 1986).  

\item {2.} The observational material used for classifying galaxies is very
often photographic plates (or copies of them). Historical reasons and the 
large angular extent of plates with respect to CCDs made this choice 
obligatory for large samples of galaxies as well as for galaxies larger than a 
few arcmin. 

\item {3.} Almost all galaxies are on the same plate or on very similar
plates exposed under similar observing conditions (seeing, sky level, etc.). 
The observational data are thus uniform, i.e. the quality of the observational 
material is the same for the whole sample. 

\item {4.} Structural components (disk, bar, etc.) are not measured
by most morphologists (Hubble's definition of types does not require
such a measurement), but only visually estimated.  Up to now, it was not
reasonable, in terms of computer time, to measure such structural
components for large samples of galaxies:  morphologists take 30 sec. to
classify a galaxy (Naim et al. 1995), whereas, for example, the 
determination of the Hubble type by means of structural components 
(see Sect. 5) may reasonably take 30 min. per galaxy. 
 
\medskip

\titlea{Problems with the traditional classification scheme}

The traditional classification method, by visual inspection of plates and by 
reference to standards for each type, has been very successful in many fields 
of extragalactic astronomy.  However this method suffers from drawbacks, 
which can become serious, depending on the use one makes of the morphological 
types.   

First, the reference to standards makes the morphological classification 
difficult enough that it resembles `more an art than a physical measurement' 
(Buta 1990).  This task is thus not accessible to most of the astronomical 
community, since, to accomplish it, one has to be an expert morphologist.  

The strong subjectivity of the task raises the question of its reproducibility 
and of the consistency of the morphological types determined by different 
morphologists.  The latter question has been addressed by Lahav et al. 
(1995).  Less than 1 \% of the large galaxies (D$_{25} >1.2$ arcmin) have 
the same morphological label when galaxies are classified in 16 bins  
by 6 well known morphologists.  The cause of the disagreement is not 
tied to differences in the images that the morphologists studied, since they 
used the same images.  The question of consistency will be discussed further 
in Sect. 4. 

The fact that structural components (bar, disk, bulge, arms, etc.), which 
astronomers naturally think of when speaking of morphological types, are not 
measured quantitatively and in some cases not even detected by traditional 
morphologists introduces differences and biaises between the presence of such 
structural components and the resulting morphological types. These differences 
and biases could be important or negligible, depending on the study 
undertaken, on the fraction of galaxies for which one or several structural 
components were missed, and on how this fraction was classified.

Second, the uniformity of the observational material is not a desirable 
property when the studied sample contains galaxies at different distances, of 
different sizes or luminosities and/or projected at different angles on the 
sky. 

\item{1.} Due to the limited dynamical range of plates, images of very 
bright galaxies are saturated and images of faint galaxies are of too low 
quality to allow any classification. As repeatedly stated by morphologists, 
this happens very often: in one third of the cases for a sample of galaxies 
larger than 1.2 arcmin (Lahav et al. 1995, Naim et al. 1995) and, more 
generally, in 85 \% of the cases (Buta 1992). This problem does not appear 
very often in the output catalogue; in other words there is no trace in the 
catalogue that some of the galaxies have classification problems. Buta (1992) 
stressed that `it is important when using published morphological types to 
know where their types came from and their limitation'.  Finally, as 
morphologists themselves admit (Lahav et al. 1995), they mostly classify 
galaxies for which they do not have suitable data. In particular for galaxies 
in nearby clusters, Dressler (1980) remarks that sky survey plates are not 
good enough for morphological classification and that Cassegrain plates (or 
prime focus plates from a large reflector) must be used. 

\item{2.} The morphological label attributed by morphologists to the observed 
galaxies unfortunately depends on the projection angle of the galaxy on the 
sky.  It is a well known fact that face-on S0 galaxies are missing in all 
catalogues of galaxies, because they are misclassified as E (van den Bergh 
1990).  The detectability of bars, arms, disks of galaxies depends strongly on 
their projection angle on the sky, as well as on the resolution of the 
observations used to perform the classification (for disks, see e.g. Nieto et 
al. 1994, whereas for bars see e.g. de Vaucouleurs \& Buta 1980 and Nieto et 
al. 1992). 

The drawback of using uniform data becomes obvious in studies of galaxies in 
more than one cluster, when the observed galaxies are not all at the same 
distance.  Andreon (1993) showed that the spiral fraction in nearby ($z<0.05$) 
clusters measured by Bahcall (1977) decreases as the redshift increases, 
reaching a null value at $z=0.05$. Such behavior (a sort of inverse 
Butcher-Oemler effect) is not intrinsic to the observed clusters, but is only 
a consequence of the increasing difficulty of identifying spiral galaxies 
as the redshift increases. A similar artificial trend has already been pointed 
out by Tammann (1987) for explaining the apparent rise of the Hubble constant 
with redshift from independent data. Such a feature is common (and has been 
overlooked) in the literature. Unfortunately, much of our knowledge of nearby 
clusters is based on such data (e.g. Sarazin 1986, Edge \& Steward 1992). 

Another problem linked to the resolution, which we discovered when studying a 
distant cluster observed with {\it HST} (Andreon, Davoust \& Heim 1996), is 
that of the sampling of the image.  The point spread functions of two images 
may have the same FWHM, but if the first image is oversampled and the other 
one has a pixel size comparable to the FWHM of the point spread function, 
morphological details will be lost in the latter image.  Thus, as one 
classifies more distant galaxies, both the resolution {\it and} the sampling 
of the image should increase. 

The use of uniform data (i.e. plates) and of images whose dynamical range is 
limited, is an easy way of collecting large numbers of morphological types, 
but it is also likely to induce misclassifications.  The images should have a 
quality adapted to the difficulty of the galaxy classification.

\titlea{Rate of agreement among traditional morphologists} 

We have no way of estimating the stability of the morphological 
classification, defined as the fraction of galaxies given the same 
morphological type when observed twice with the same observational material by 
the same morphologist. However we can estimate the effect of the subjectivity 
of the Hubble types' definition and how it is related to the quality of the 
observational material by measuring the reproducibility of the Hubble type 
estimate in the cases when only the morphologists differ, when only 
observations differ and when both differ. 

\begtabfull
\tabcap{1}{Agreement on the Hubble type estimates}
\halign{#\hfill&\quad\hfill#\hfill\cr
\noalign{\hrule\medskip} 
\multispan2{\hskip 1.5truecm Different morphologists\hfill}\cr
Large galaxies& 40 \% - 85 \% \cr
{\it HST} images of distant clusters& $\sim$80 \% \cr
\noalign{\medskip} 
\multispan2{\hskip 1.5truecm Different images\hfill}\cr
Large galaxies& $\sim$80 \% \cr
{\it HST} images of distant clusters&  $\sim$80 \% \cr
\noalign{\medskip} 
\multispan2{\hskip 1.5truecm Different images and morphologists\hfill}\cr
Coma galaxies: Dressler {\it vs} Butcher \& Oemler& 88 \% \cr
Coma galaxies: Dressler {\it vs} Rood \& Baum& 84 \% \cr
Coma galaxies: Dressler {\it vs} RC3& 73 \% \cr
\noalign{\medskip\hrule}
}
\endtab 

\titleb{Different morphologists}

We start with the results of a comparative classification exercise 
performed among expert morphologists using the same images (Lahav et al. 1995, 
Naim et al. 1995). Six expert morphologists classified 835 (nearby and 
large) galaxies in 16 classes.  All the morphologists looked at exactly the 
same laser printed images, except for one, who looked at images on a computer 
screen. 

The solid line in the left panel of Fig. 1 shows the relative agreement
(in \%) among morphologists (grouped by pairs) on the morphological type of a 
given galaxy.  This agreement is measured by the fraction of galaxies
(among 835) given the same coarse Hubble type (E, S0, or S+Irr) by a pair of 
morphologists.  We put the galaxies in one of the three bins according to 
the T value listed in Naim et al. (1995).  
The fraction of galaxies given the same morphological type 
ranges from $\sim 40$ \% to $\sim 85$ \%, depending on the pair of 
morphologists, with a mean of $\sim 50$ \%.  

The dotted line in the same panel shows the relative agreement for easily 
classified galaxies (i.e. for galaxies whose morphological type is not 
followed by a colon or question mark).  This is the fraction of galaxies 
(among the ones easily classified by both morphologists of a pair) that were 
given the same coarse Hubble type by both morphologists.
The agreement is much better, of the 
order of 90 \%, which is normal, since the task is admittedly easy.

The right panel of Fig.1 shows that this good agreement in fact only concerns 
a minor fraction, 50 \% and often less, of the sample.  It does not show how 
often pairs of morphologists agree on the type of a galaxy, but how often 
they agree on whether the galaxy is easy to classify or not.  This fraction 
of galaxies which are easily classified by pairs of morphologists (i.e. whose 
type is not followed by a colon or question mark for either morphologist of the 
pair) is rather small. On this small fraction the agreement on the type is 
excellent.

Furthermore, part of the agreement among morphologists is due to chance, since 
we have reduced the number of bins from 16 to 3.

\begtabfullwid
\centerline{\psfig{figure=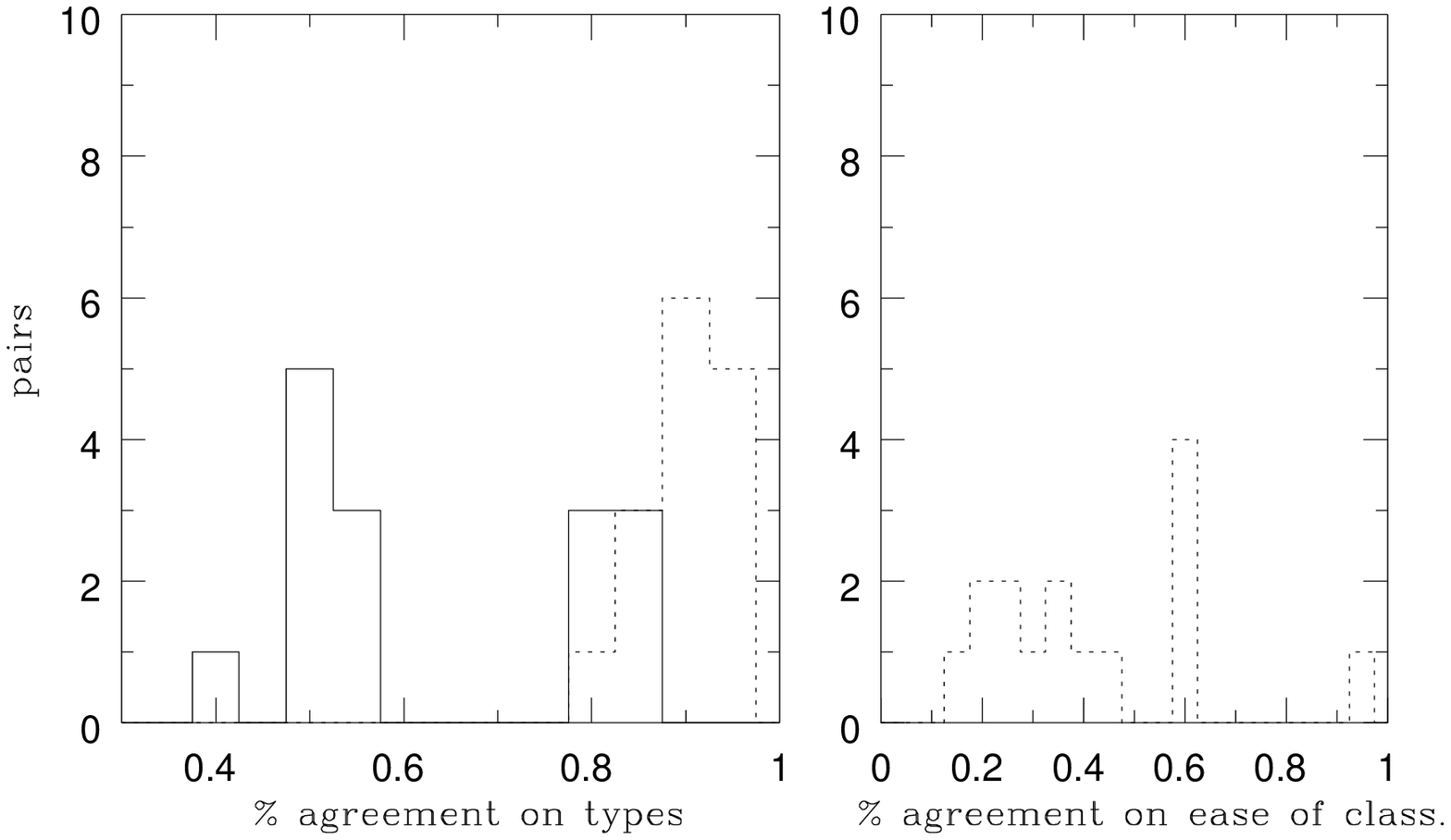,width=15truecm,bbllx=35mm,bblly=125mm,bburx=210mm,bbury=240mm}}
\figure{1}{Left panel: relative agreement among pairs of morphologists on the 
morphological types, for the whole sample of galaxies (solid line) and for 
easily classified galaxies only (dotted line). Right panel: relative agreement 
among pairs of morphologists on whether a galaxy is easy to classify.} 
\endtab

Dressler and Oemler classified the galaxies of the Abell 851 cluster from the 
same {\it HST} images before refurbishing (Dressler et al. 1994a). Couch, 
Sharples and Smail did the same for the galaxies of Abell 370 (Couch et al. 
1994).  In both cases, the disagreement among morphologists on the assignment 
of the morphological type was about 20 to 25\%.                                 

\titleb{Different images}

\medskip

We now compare the estimates of a morphologist looking at different images. 
Dressler et al. (1994b) classified twice the galaxies of the 
distant cluster Cl 0939+4713 observed with {\it HST}, before 
and after refurbishing. Taking the types of galaxies from their Fig. 4, 
we find that 20 \% of the galaxies have different coarse Hubble types.  

Naim et al. (1995) and we find a similar rate of disagreement between the 
morphological types assigned by de Vaucouleurs (Naim et al. 1995) and by RC3 
(de Vaucouleurs et al. 1991) for the sample of 835 galaxies. 

\titleb{Different morphologists and images}

The agreement, when both morphologists and images differ, can be estimated 
from the published morphological types of galaxies in the Coma cluster. We 
take Dressler's (1980) morphological type as reference, because it concerns a 
larger and deeper sample than the others.  The agreement between Dressler 
(1980) on the one hand, and other morphologists (Rood \& Baum 1967, Butcher \& 
Oemler 1985) and RC3 (de Vaucouleurs et al. 1991) on the other, on the types 
of Coma galaxies is shown in the last three rows of Table 1. The agreement, 
for classifications based on different observational materials, is relatively 
good, even if our comparison is biased by the fact that low values of the 
agreement are not permitted because of the small number of classes which 
naturally induces some chance agreements. We have checked that the chance 
agreement can be as large as 35 \% to 75 \% depending on the morphological 
composition of the sample and on the exact way in which we randomly classify 
the galaxies (conserving or not the morphological composition of the sample).  

\medskip
These three comparisons, when only morphologists differ, when only 
images differ and when both differ, show that about 20 \% of the 
galaxies have an unprecise Hubble type.  This inaccuracy arises because of the 
quality of the material used for the morphological estimation and/or because 
of the subjective nature of morphological classification. The fraction 
is even larger if galaxies are classified from Schmidt plates. 

\titlea{Classifying by means of structural components}

In view of the drawbacks of the traditional method, especially when applied to 
the specific task of classifying galaxies in clusters, other methods have been 
introduced. Most of them have the advantage of giving a stable estimate of the 
Hubble type, in particular when prototypes or training sets are kept fixed, 
but they ultimately infer the Hubble type from the galaxy properties by using 
the fact that the Hubble type is correlated to the galaxy properties (colors, 
bulge on disk ratio, spectral shape, etc.). The observed properties of 
galaxies in classes defined in this way are strongly influenced by the 
constraints used to define the classes; for example, if colors are used for 
classifying galaxies, blue ellipticals will be missing whether they exist or 
not in the Universe. We are thus not in favor of adopting such methods. 

We propose using an alternative classification method based on the
criteria presented in Michard \& Marchal (1993) and adopted with minor
changes in Andreon et al. (1996).  This method relies on the detection of
structural components, rather than on the resemblance to standards. 
 
The morphological classification is done as follows.  A number of radial 
profiles (of surface brightness along the major and minor axes, of 
ellipticity, of the position angle, of isophote deviations from the perfect 
elliptical shape) are computed and then visually inspected for the detection 
of signatures typical of the morphological components described below.  When a 
component (a disk, a bar, an envelope, etc.) is detected, we take note of it. 

The segregation between the E and S0 types is based upon the examination of 
the surface brightness profile along the major axis of the projected galaxy, 
plotted on an r$^{1/4}$ scale. The presence of a disk gives a characteristic 
bump above the linear profile which characterizes pure spheroids. This 
photometric signature of S0s should not be confused with other often observed 
deviations from the de Vaucouleurs law, both in giant ellipticals with an 
envelope enhanced above the extrapolated r$^{1/4}$ line, and in minor objects 
with evidence for a cut-off below this line (Schombert 1986).  This signature may be absent in 
some face-on S0s, which are then classified as Es; there is no way of avoiding 
this type of misclassification, as shown by the photometric similarity between 
the face-on NGC 3379, prototype of Es, and the edge-on NGC 3115, prototype of 
S0s (Capaccioli et al. 1991). Round galaxies with elliptical isophotes and 
exponential profiles are classified as S0s. In this spirit, these criteria are 
identical to those used in classical morphology, when the observer looks for 
subtle changes of gradient in a galaxian image. 

The E galaxies are further subclassified into `disky', or diE, `boxy', or
boE, and `undefined', or unE. This is done mainly from the radial profile
and sign of Carter's $e_4$ coefficient\FOOTNOTE{The $e_4$ coefficient
characterizes the deviations of period $\pi/2$ and phase 0 of the isophote
shape from the perfect ellipse. Galaxies with positive $e_4$ show a
luminosity excess along the major axis interpreted, as for S0 galaxies, as
revealing the presence of a disk.  A negative $e_4$ coefficient indicates
boxy isophotes.}.  Roundish galaxies where the $e_4$ values fluctuate
around zero are classified unE. To this class also belong Es with inner
disky and outer boxy isophotes (or vice versa), and lemon-shaped Es.  Note
however that non zero $e_4$ are also produced by the presence of dust and
shells, which also give a signature in higher-order coefficients and in
the two-dimensional image. 

The S0s are subclassified into the SA0, SAB0 and SB0 families, based upon
the radial profile of the position angle (PA) of the isophotal major axis. For 
a galaxy observed with a sufficient resolution, the PA of the bulge and disk 
are nearly the same, while the PA of the bar stays in a limited range 
(except in cases of unfavorable projection). At the distance of nearby 
clusters ($v \sim 5000-7500$ km s$^{-1}$), it often happens that the contrast 
between bulge and bar is washed out by the seeing. 

The segregation between S0 (or E) and S is based on the presence of
spiral arms or very irregular or asymmetric isophotes. Because of this
choice, the spiral class defined in such a way may include (a small number of)
galaxies showing strong signs of interaction (tidal arms, double nuclei,
non concentric isophotes, etc.).  On the one hand, such a misclassification
could be viewed as a contamination of the `pure' spiral class; on the other
hand, interacting galaxies often show important star formation. 
But in no case should we introduce another class, that of interacting 
galaxies, because this goes one step beyond that of classification, since we 
then interpret peculiarities in terms of gravitational interaction.

Spiral galaxies are not classified in more subclasses than S and Irr for
two reasons. First of all, we lack an objective and unique method to
distinguish stages in the S class. As noted by Sandage (1961), Hubble's
(1936) description of spirals does not allow a unique classification of
some spirals (e.g. NGC 4941) and some criteria have to be relaxed
(Sandage, 1961, p. 13).  Furthermore, splitting the S into the traditional
stages does not separate galaxies of different physical types, since each
stage is a blend of galaxies showing a range in properties, sometimes as
large as the one found among the stages (see, e.g., Gavazzi \& Trinchieri
1989, Roberts \& Haines 1994, and Staveley-Smith \& Davies 1988). 

Finally, we stress that the structural classification scheme does not 
(implicitly or explicitly) use any other structural parameter, different from 
the ones above, to classify the galaxies; in particular, it does not make use 
of the bulge to disk ratio to discriminate between lenticulars and spirals. 

In summary, the galaxy morphological type is nothing else than the obvious
composition of morphological components detected following the rules
outlined above. For example, a galaxy having (almost perfect) elliptical
isophotes with a bump in surface brightness along the major axis and a
characteristic twist of the position angle, is noted as having a
photometric disk, a bar and no spiral arms, and therefore is classified as
SB0.  For this galaxy, whether it is boxy or disky is not a matter of
concern, since boxiness or diskiness have no effect on its morphological
type. 

The Hubble sequence therefore looks like this :

\medskip
\halign{#\hfill&\quad\hfill#\hfill&\quad\hfill#\hfill&\quad#\hfill\cr
\noalign{\hrule\medskip} 
detailed types: & boE unE diE & SA0 SB0 & S Irr \cr
coarse Hubble types:  & --------E-------- & -----S0----- & ---S--- \cr
\noalign{\medskip\hrule}} 
\medskip

The criteria of structural morphology are as close as possible to 
those of morphologists.  The main difference is that structural components are 
measured, and galaxies are classified according the presence (or absence) 
of these structural components {\it without exceptions} and that
galaxies are not classified according to the resemblance to standards.
This alternative approach has several advantages.

\titlea{Advantages of the structural classification scheme} 

\titleb{Low subjectivity}

In the first step of the classification work, where we compute the various 
profiles, the only (very minor) subjective task is to select foreground or 
background objects to be masked because they contaminate the galaxy 
brightness. 

The second step, the visual detection of the morphological components, is 
still partly subjective, in particular when we are classifying galaxies on the 
borderline between classes.  But nothing prevents us from making this part of 
the analysis automatic, by introducing profile templates which mark the 
expected behavior for the morphological components, thus making it fully 
objective and reproducible.  However, the complexity of Nature, which produces 
galaxies with disks, halos, bulges and bars with a variety of ellipticity and 
surface brightness profiles, and which mixes them and shows them to us 
projected on the sky under different angles, makes this {\it automatic} 
approach too complex just for determining the morphological type. 

The third step of the classification method, assembling the structural 
components to derive the morphological type, is of course fully automatic.

The subjectivity arising from the visual detection of morphological components 
in our method is always lesser than in the traditional estimation of the 
types, since the detection of morphological components is easier on profiles 
than on direct images. Therefore, the number of difficult cases of 
classification is reduced. For example, deviations from de Vaucouleurs' law 
are easily visible on a surface brightness vs. $r^{1/4}$ profile, used by us, 
much more so than on direct galaxy images, used by classical morphologists. 
This allows us to detect disks in many almost face on S0s. In fact, the 
ellipticity distribution of Dressler's (1980) S0s in Coma shows a clear bias 
against round (face-on) galaxies (Michard 1996). Using our criteria to 
classify the galaxies, 1/3 of Dressler's (1980) Es move into the S0 class, 
thus strongly reducing the bias (Michard 1996). 

The residual role played by the subjectivity can be estimated by comparing the 
rate of agreement among Hubble types estimated by different authors from the 
same galaxy profiles. 

\titleb{Reproducibility}

A more objective method should produce highly reproducible results.
To check the reproducibility, and to understand the effect of any personal 
judgment on the morphological estimate, (at least) two authors  classified a 
large subsample of galaxies in Coma independently, after a period of training. 
The galaxies to be classified had magnitudes in the range 12 to 17 B mag and 
their images had a typical seeing in the range 0.35 to 3 arcsec (see Andreon 
et al. 1996 for details). We had both CCD and digitized plate images available 
for classification.  For this comparison, the images considered and all 
derived data (profiles of surface brightness, ellipticity, position angle, 
$e_4$, etc.) were exactly the same.  

A perfect agreement on the coarse Hubble type was found for all (more 
than 100), but two, galaxies.  The two discrepant galaxies are peculiar: a 
very dusty galaxy (GMP 1646) with an r$^{1/4}$ profile (elliptical or 
irregular?) and a dusty asymmetric lenticular (or spiral?) (GMP 1204). 

The fact that, in the traditional morphological analysis, 15-20\% of the 
galaxies have different morphological estimates introduces a scatter in the 
properties of the Hubble types {\it precisely because of the highly subjective 
nature of the type estimate} and this can mask real differences between the 
properties of the morphological types. With the reproducibility of 95\% 
reached by our estimate of the type, only a small fraction of galaxies are not 
of the type assigned by us; this greatly reduces the scatter in the properties 
within classes and allows an easier comparison of the types in different 
locations in the Universe (cluster vs field, nearby vs distant, etc.) and 
greatly helps detecting previously unnoticed properties of the morphological 
types. 

\titleb{Stability with respect to observing conditions}
 
To understand whether different telescope set-ups, seeing conditions, filters, 
and other effects affect our morphological classification, we observed the 
same subsample of galaxies in Coma during different runs, at different 
telescopes, with different filters (mainly Johnson V and Gunn r) and detectors 
(4 CCDs and one plate).  Different images of the same galaxy were classified 
by the same or by different observers. Once the observing conditions are 
taken into account, there is perfect agreement on the coarse Hubble type and 
on the detailed type of the galaxies for all 54 comparisons done, 
but for GMP 1300, classified as S from our plate and  S0 from our CCD image.  
Typical differences tied to observing conditions are due, for example, to the 
limited field of view of observations at the 3.6m Canada-France-Hawaii 
telescope in Hawaii (CFH), preventing one from classifying galaxies larger 
than the CCD field of view, or to the lower resolution of the prime focus 
plate used (FWHM=1.8\arcsec) compared to the best quality CCD images. The 
comparison of the CFH observations taken under excellent seeing conditions 
(FWHM=0.35\arcsec) with observations at the 2m telescope of the Pic du Midi 
Observatory taken under good to fair seeing conditions 
(0.8\arcsec$<$FWHM$<$1.5\arcsec) for common galaxies shows that Pic du Midi 
observations are good enough for determining the galaxy type, since not one 
galaxy (out of 13) was classified in two different coarse classes 
from the two observing materials (again, once the small field of view of CFH 
observations is taken into account).  The CFH images certainly allow one to 
classify E galaxies more easily into the three families and to detect small 
morphological components, such as dust lanes. 

\titleb{Ability to bring out new properties of galaxies in clusters}

Our morphological scheme, {\it whatever it traces}, gives highly reproducible 
results and is not very sensitive to the observing conditions, provided the 
resolution is adapted to the difficulty of the galaxy classification. 
But this does not necessarily mean that our method brings galaxies into classes 
containing objects with similar physical properties better than the 
traditional classification scheme. 

The {\it relative} quality of our classification method can only be judged on 
its results, i.e. on the ability of our scheme to separate galaxies of 
different {\it physical} types into different classes, and to bring out new 
properties of galaxies in clusters.  

This ability is demonstrated by a series of recent results obtained with this 
method. For example, we have detected a segregation of the morphological types 
{\it stronger} than the usual clustercentric or density segregations in the 
Perseus (Andreon 1994) and Coma (Andreon 1996) clusters and in the distant 
cluster Cl0939+4713 (Andreon, Davoust \& Heim 1996). Es have a fainter mean 
surface brightness than S0s in Coma (Andreon 1996) and Cl0939+4713 (Andreon, 
Davoust, Heim 1996). Furthermore, by using images of resolution adapted to the 
difficulty of classification, we have found that the spiral fraction rises by 
a factor 2 or 3 in the Perseus and Coma clusters (Andreon 1994, 1996) and, by 
indirect evidence, in most nearby clusters (Andreon 1993), but not in the 
distant cluster Cl0939+4713; this strongly reduces the evidence for a 
morphological evolution of galaxies in clusters that many observations make 
unreasonable (see Andreon 1993, and Andreon, Davoust \& Heim 1996 for 
details). 

\titlea{Comparison between structural types and traditional ones} 

In order to assess the relation between our morphological types and the 
`traditional' ones, we now present a detailed 
comparison between our types and published ones for galaxies in the Coma 
cluster. 
Because of the scarceness of Es and S0s classified into detailed classes 
(diE/boE, SA0/SB0), mainly in the comparison samples, but also in ours, 
we can only estimate the quality of the broad Hubble types (E,S0,S). 

This comparison of Hubble type estimates is based on a magnitude complete 
sample of 190 galaxies in Coma brighter than mag$_B = 16.5$ mag and within one 
degree from the cluster center. Types determined by structural morphology are 
presented in Andreon et al. (1996) and Andreon et al. (in preparation), 
traditional types are taken from the literature.  In summary, our structural 
classification is based on (digitized) Schmidt plates for obvious spirals and 
on the `quantitative analysis' of a digitized KPNO 4m prime focus plate, of 
CCD images taken under good or excellent seeing conditions at Pic du Midi and 
CFH for all the other types (as well as for non obvious spirals). Significant 
overlap exists among our observations, allowing us to assess the relative 
quality of the classifications (Sect. 6.3). For the classifications from the 
literature, particular attention was paid to the meaning given to the type 
notation by each morphologist: for example, the S0/a type is the intermediate 
type between S0 and Sa types for some authors {\it and} a sign of the 
inability to discriminate between the two types for others. 

\titleb{Saglia, Bender \& Dressler (1993)}

Saglia, Bender \& Dressler (1993, hereafter SBD) determined the coarse Hubble 
type of galaxies in Coma following a morphological scheme similar to ours, in 
the sense that structural components of galaxies are measured, and not visually 
estimated. They take however a conservative approach, changing the traditional 
type of the galaxies only when their analysis shows that the type listed in 
Dressler (1980) is wrong.  

Table 2 shows the results
of the comparison of SBD's Hubble types with ours for common galaxies.

The fraction of discrepant types is 15 \% (7 galaxies out of 47).

-- One galaxy (GMP  1201) was classified as S0 by us (and by all the other 
morphologists, including Dressler) and S by SBD. 

-- Out of the two galaxies classified E by SBD and S0 by us, one (GMP 1878) 
shows a faint but extended disk, and is certainly not a boE, as our 
0.35\arcsec\ resolution CFH images show, and the other one was assigned the 
same type by us from the analysis of two independent images. All the other 
galaxies with discrepant types are faint (mag$_B >16.0$) and have low contrast 
spiral arms, and were thus classified as S0/a by us and S0 by SBD. 

The fraction of discrepant types (15\%) is very low, and fully understood
but for two galaxies out of 47 (4 \% of the sample).

\begtabfull
\tabcap{2}{SBD vs us}
\halign{#\hfill&\quad\hfill#&\quad\hfill#&\quad\hfill#\cr
\noalign{\hrule\medskip} 
SBD /us     &   E    &   S0   &   S  \cr
\noalign{\medskip\hrule\medskip} 
E & 19 & 2  & 0 \cr
S0 &  0 & 18 & 4 \cr
S+I &  0 &  1 & 3 \cr
\noalign{\medskip\hrule}}\endtab

\titleb{Butcher \& Oemler (1985)}

Butcher \& Oemler (1985, hereafter BO) classified galaxies in the Coma cluster 
core by inspection of a Schmidt plate (scale $\sim 65$ arcsec mm$^{-1}$) 
and/or of an unspecified ``4m telescope" plate. Their classification scheme has 
a fine resolution for bright and large galaxies, and a coarse one for faint 
(mag$_J>15.5$) galaxies and for galaxies difficult to classify (e.g. face-on 
S0s). At their intermediate resolution, which corresponds to our coarse 
Hubble types, most of the galaxies have been assigned a Hubble type by BO,
whereas a small percentage could not be put in a single class by BO (some 
galaxies that are noted as EL, i.e. E or S0). BO put the S0/a galaxies in the 
S bin in the intermediate and coarse resolution schemes (see their Table 12). 
Since we use their intermediate resolution, we put our S0/a in the S bin for 
the purpose of comparison.  

Table 3 shows the results of the comparison of the morphological types for 
common galaxies. Out of 123 galaxies in common, 23 (=19 \%) have discrepant 
Hubble types. 

-- Out of the 8 early-type galaxies with discrepant types, four (GMP  999, 
1035, 694, 552) are faint (mag$_B \sim$ 16), two (GMP  1373 and 908) have an 
uncertain type in our work, one (GMP  565) has a disk but not bright enough to 
allow one to classify it as S0, in spite of BO's S0 classification, and 
finally our data for the last galaxy (GMP  1834), classified as E by BO, 
suggest the presence of a small bar (undetected by BO) which lead us to 
classify this galaxy as SAB0. 

-- Out of the 15 galaxies classified as S0 by BO and as S by us, 7 show a 
spiral pattern not resolved into spiral arms and without HII regions on our CCD 
images. Such faint features were probably not detected on BO's plates, or 
judged of null importance by them for the classification. 

The galaxies with discrepant types are not distributed uniformly among 
the Hubble types. As a consequence, the morphological composition of the 
cluster differs in the two works, {\it even though} the sample is composed 
of the same galaxies.  The spiral fraction (here defined as 
the ratio of the number of spirals to the total number of galaxies) in Coma's 
core rises from 17 \%, when computed with BO's morphological types, to 26 \% 
when computed with our types. 

\begtabfull
\tabcap{3}{Butcher \& Oemler vs us}
\halign{#\hfill&\quad\hfill#&\quad\hfill#&\quad\hfill#\cr
\noalign{\hrule\medskip} 
     BO /us     &   E    &   S0   &   S  \cr
\noalign{\medskip\hrule\medskip}
       E & 23 & 3  & 0  \cr
      S0 &  5 & 48 & 15 \cr
       S &  0 &  0 & 20 \cr
n.c. (EL)&  3 &  6 & 0 \cr        
\noalign{\medskip\hrule}}\endtab

\begtabfull
\tabcap{4}{Dressler vs us}     
\halign{#\hfill&\quad\hfill#&\quad\hfill#&\quad\hfill#\cr
\noalign{\hrule\medskip} 
      Dressler /us     &   E    &   S0   &   S \cr
\noalign{\medskip\hrule\medskip}
       E+D+E/S0 & 31 & 12 & 0   \cr
      S0+S0/a &  9 & 58 & 11  \cr
       S+I &  0 &  3 & 29  \cr
\noalign{\medskip\hrule}}\endtab

\titleb{Dressler (1980)}

Dressler (1980) classified the Coma galaxies in a slightly larger region than
ours, by inspection of a plate taken at the Cassegrain focus of a 2.5m
telescope (scale $\sim 10.9$ arcsec mm$^{-1}$).  His sample is the largest one 
in the literature for Coma galaxies.  It is part of the largest published 
survey of morphological types of galaxies in nearby clusters. The plates used 
by Dressler have one of the smallest scales, and therefore meet our first 
requirement, to be of good quality.  Dressler's types have been widely used in 
the literature for studies of morphological segregation (e.g. Whitmore \& 
Gilmore 1991, Whitmore et al. 1993, Sanrom\`a \& Salvador-Sol\'e 1990), and 
most of our knowledge on morphological segregation of galaxies in clusters 
rests on data listed in this catalogue.  Dressler's X/Y notation for types 
points out his care in not definitely classifying galaxies for which he does 
not have  suitable data (Dressler 1980). Therefore Dressler's X/Y types are not 
meant to be a transition type between X and Y. 

Table 4 shows the comparison of types for common galaxies. We put Dressler's 
E/S0 and D types in the same bin as the E. As for previous comparisons, we put 
Dressler's S0/a galaxies and ours in the S class,
but consider them separately in the following discussion. 
The fraction of discrepant types is 23 \% (35 galaxies out of 153).

-- 9 of Dressler's S0s were classified as E by us;
most of them are roundish galaxies difficult to classify because face on. 

-- 12 of Dressler's E (+D+E/S0) were classified as S0 by us, of which 6 are 
faint galaxies, 5 were classified by Dressler as E/S0 (i.e. E {\it or} S0) and 
the last one (GMP  1931) is a barred galaxy. 

-- 11 of Dressler's S0 (+S0/a) were classified as S by us.  The fact that 
at least 7 of them just have a spiral pattern not resolved into arms and HII
regions in our images suggests that low contrast spiral arms are not 
detected or are considered of null importance in Dressler's morphological 
scheme. 

-- Two of Dressler's Sa galaxies (GMP 1925 and GMP 1844) and one S0/a (GMP 
1154) were classified as S0 by us. Dressler's S0/a (i.e. S0 or Sa), classified 
S0 by us, is not a discrepant case since Dressler's classification is 
uncertain. Furthermore our images of this galaxy show a slight asymmetry that 
may explain why Dressler's classification is uncertain. The two other galaxies 
show uncommon characteristics of lenticulars, an isophotal twist in the outer 
envelope (GMP 1925) and an important ring-lens (GMP 1844).  It is possible 
that Dressler used such characteristics to classify the galaxies as S, because 
they do not present spiral arms or dust on our images. In such a case, the 
differences in the morphological estimates are due to the different weights 
given to structural components in the definition of the galaxy types. 

The fraction of discrepant types is 23 \%, (at least) half of which are 
accounted for by galaxies difficult to classify, and the other half by 
different weights given to the presence of a spiral pattern or other 
morphological structures in classifying galaxies. 

\titleb{Rood \& Baum (1967)}

Rood \& Baum (1967, hereafter RB) classified the galaxies in the Coma cluster
by visual inspection of a prime focus plate from the 5m Palomar telescope
(scale $\sim 27$ arcsec mm$^{-1}$). 

Table 5 shows the comparison of RB's morphological types and ours for galaxies 
in common. We put the S0/a galaxies in the S bin as in previous comparisons, 
and consider them separately in the following discussion. 

-- Five of the 6 early-type galaxies with discrepant types are faint. The last 
one, classified E by us and S0 by RB, is the second ranked galaxy of the Coma 
cluster.  RB probably classified it as S0 because it has an outer envelope, 
which identifies this galaxy as an S0 according to Hubble's definition of S0. 

-- All the other discrepant types concern RB's S0s classified as S by us.  
Since RB used similar plates to ours (with similar scales), they probably 
did not miss the structural components that we detected on these galaxies, but 
they presumably judged these components of null importance for the 
classification. We note, furthermore, that the missed spirals do not have 
obvious spiral arms or HII regions but just a spiral pattern. From the 
classical morphologist's point of view, these galaxies do not resemble the 
standard Sa, which explains their classification as lenticulars. 
                                         
The fraction of galaxies with a discrepant type is 16 \%, if we put
the S0/a in the S class, or 12 \% if we put the S0/a in the S0 class, as 
RB seem to do. 

\begtabfull
\tabcap{5}{Rood and Baum vs us}     
\halign{#\hfill&\quad\hfill#&\quad\hfill#&\quad\hfill#\cr
\noalign{\hrule\medskip} 
       RB /us     &   E    &   S0   &   S \cr
\noalign{\medskip\hrule\medskip}
       E & 20 & 5  & 0   \cr
      S0 &  1 & 24 & 10 \cr
       S &  0 &  0 & 5 \cr  
\noalign{\medskip\hrule}}\endtab

\begtabfull
\tabcap{6}{RC3 (all RC3 Coma galaxies) vs us}     
\halign{#\hfill&\quad\hfill#&\quad\hfill#&\quad\hfill#\cr
\noalign{\hrule\medskip}
       RC3 /us &   E    &   S0   &   S \cr
\noalign{\medskip\hrule\medskip}
       E & 22 & 5  & 0  \cr
      S0 &  7 & 17 & 9  \cr
       S &  1 &  2 & 16 \cr
    n.c. & 13 & 39 & 19  \cr      
\noalign{\medskip\hrule}}\endtab

\begtabfull
\tabcap{7}{RC3 (bright RC3 Coma galaxies) vs us}     
\halign{#\hfill&\quad\hfill#&\quad\hfill#&\quad\hfill#\cr
       RC3 /us     &   E    &   S0   &   S \cr
\noalign{\hrule\medskip}
       E & 18 & 2 & 0  \cr
      S0 &  6 & 10 & 6  \cr
       S &  1 & 1 & 9  \cr
    n.c. &  1 & 3 & 3  \cr          
\noalign{\medskip\hrule}}\endtab

\titleb{RC3}

The RC3 catalogue (de Vaucouleurs et al. 1991) lists a large number of 
galaxies (more than 25000) distributed over the whole sky together with their 
morphological type.  It is not fair to request that this huge catalogue be as 
good in selected regions as detailed studies of these regions.  Furtermore, the 
galaxy types are not all based on the same observational material, but on a 
variety of plates, films, prints or copy of them, and have been assigned by 
different morphologists. 

Because RC3 is the main catalogue of reference for types, and because 
morphological types are often claimed good if they agree with the ones 
listed in RC3 (e.g. Doi, Fukugita \& Okamura 1993), 
we have to investigate how our types are related to 
the RC3 ones. A major caveat is in order: Coma galaxies are fainter than the 
majority of galaxies listed in RC3, and, as a consequence, 
we are making a comparison with the faint end of the catalogue only. 

Among the 190 galaxies of our complete sample of galaxies in Coma, 129 are 
listed in RC3. Among those, only 71 have been assigned a 
definite morphological type in RC3, whereas 14 others have an uncertain 
morphological type. Limiting our sample to the magnitude of the faintest 
galaxy classified in RC3 (mag$_B=16.3$), 79 brighter galaxies (out of 158) 
are not classified (or listed at all) in RC3. This catalogue is 
therefore incomplete to a large degree for such faint galaxies. To reach a 
reasonable degree of completeness (90 \%), RC3 has to be limited 
to mag$_B = 15.45$, and in this case the sample of classified galaxies 
only numbers 53 galaxies. 

Tables 6 and 7 show the comparison between our types and the ones listed in 
RC3 for the whole sample (mag$_B<16.3$) and for a brighter sample 
(mag$_B<15.45$) for which RC3 is 90 \% complete. First of all, 60 \% of the 
S0s are missing in RC3 whereas only 30 \% of Es and 43 \% of Ss are missing. 
The missing galaxies are therefore not distributed uniformly among types. We 
stress again that these missing galaxies are brighter than the chosen 
magnitude limit. 

The fraction of galaxies with discrepant types is 30 \% for the two RC3 
subsamples, a factor two higher than the value found for studies dedicated to 
the Coma cluster.  A similar disagreement is found between RC3 and Dressler 
(27 \%), based on 66 common galaxies. The disagreement between RC3 and SBD is 
similar to the one between us and BO, Dressler or RB (i.e. $\sim 15$ \%). 

RC3 agrees extraordinary well with BO and RB (less than 5 \% of 
disagreement). This is inconceivable for two reasons. First, no pair of 
morphologists shows such a good agreement, and never does the RC3 type agree 
so well with de Vaucouleurs' estimate of the morphological type reported in 
Naim et al. (1995).  Second, RC3 is a collection of types estimated by 
different morphologists. The only explanation for such a good agreement is 
that the morphological types of Coma galaxies in RC3 were taken from BO and 
RB. 

In summary, RC3 is incomplete to a large degree in this region and for 
such faint galaxies, and, moreover, the missing galaxies are not distributed 
in a uniform way among types. Finally, one third of the galaxies with a 
morphological type in RC3 have discrepant types with respect to ours and 
Dressler's. Therefore, the use of this catalogue for morphological studies of 
nearby clusters is not recommended. Moreover, due to the high 
misclassification of the RC3 Coma sample of galaxies, this sample is not a 
good comparison sample to estimate the quality of other morphological 
classifications, as has unfortunately been done in the literature. 

\titlea{Redshift dependence of the classification}

Up to now, we have compared our morphological types with those of traditional 
morphologists for {\it present day} galaxies.  But 
how does the quality of our type estimates depend on redshift? 
Since the redshift does no enter at all in the classification skill, our 
expectation is that the quality of our Hubble estimates is as good for distant 
as for nearby galaxies, provided, of course, that the images of distant 
galaxies are as good as the nearby ones in terms of {\it rest-frame} 
resolution, sampling, depth, etc. 

Distant ($z>0.3$) clusters are now currently observed by {\it HST} and their 
galaxies are beginning to be classified by traditional morphologists.  
Dressler \& Oemler classified the galaxies of the {\it distant} cluster 
Cl0939+4713 ($z\sim0.4$) (Dressler et al. 1994b). For 31 galaxies, their type 
is listed in Stanford, Eisenhardt \& Dickinson (1995). We re-classified the 
same galaxies from the same images (Andreon, Davoust \& Heim, 1996). For these 
galaxies, we found the same rate of agreement (23 \%) between our and Dressler 
\& Oemler's classifications as for {\it nearby} galaxies. This result implies 
that the agreement between our classes and the traditional ones is not 
strongly redshift-dependent and that Dressler \& Oemler classify distant 
galaxies in the same way as they classify nearby ones. 

\titlea{Conclusion} 

When our structural morphological scheme is reduced to the 3 coarse Hubble 
types and is applied to a sample of galaxies in the Coma cluster, it agrees to 
within 15 or 20\% with other detailed traditional morphological analyses. 
This agreement is comparable to the one obtained among traditional 
morphologists. This shows that our criteria for classifying galaxies 
lead to the same results as the standard ones, and that our method is 
acceptable for classifying galaxies. 

Most disagreements occur for galaxies difficult to classify because they 
are faint (mag$_B > 16.0$), because images of higher resolution are 
required, and/or because of a disagreement on the significance of structural 
parameters in borderline cases between S0 and S.

At least half of the disagreements arise in spiral galaxies whose spiral 
arms (or spiral pattern) have not been detected or taken into account by 
previous analyses, even if S0/a galaxies are spirals by definition (e.g. BO, 
RC3 p. 15). Therefore traditional morphologists underestimate the spiral 
fraction in nearby clusters by 7 to 10 \%.  Even though this does not seem to 
be a large quantity, it represents an error of 100 \% in the relative number 
of spirals in clusters, because of their scarcity with respect to early-type 
galaxies. This helps to resolve apparent differences between the number of 
spirals in nearby and distant clusters, for which various explanations and 
theories have been proposed in the recent past.  At any rate, this underscores 
the need for higher resolution images as one classifies fainter and/or smaller 
galaxies. 

The method we adopt uses quantitative criteria (the presence of structural 
properties) to classify galaxies, our types are thus reproducible and stable 
to a higher degree than the ones of morphologists.  We also use better 
observational material for galaxies difficult to classify.  The first results 
of analyses based on this method, summarized in Sect. 6.4, are promising.
For all these reasons, our classification method should be preferred for 
classifying galaxies in clusters, even if it is more time and telescope 
consuming than the others. 

\acknow{SA warmly thanks  R. Michard for his wise suggestions on morphological 
classification, and Prof. G. Chincarini, whose scepticism prompted us to write 
this paper. We thank Dr. A. Naim for giving us the morphological type of the 
galaxies in his sample in a computer readable form. We also thank Chris Moss 
for convincing of the importance of understanding our results on morphological 
segregation.  We thank G. Soucail for information concerning typical 
observations of distant clusters with {\it HST}.} 

\begref{References}

\ref Andreon S., 1993, A\&A 276, L17

\ref Andreon S., 1994, A\&A 284, 801

\ref Andreon S., 1996, A\&A in press

\ref Andreon S., Davoust E., Heim, T., 1996, A\&A, submitted

\ref Andreon S., Davoust E., Michard R., Nieto J.-L., Poulain P. 1996, A\&AS, 
116, 429 

\ref Bahcall N., 1977, ApJ 218, L93

\ref Buta R., 1990, in {\it Morphological and physical
classification of galaxies}, ed. G. Longo, M. Capaccioli and
G. Busarello, p. 1

\ref
Buta R., 1992, in {\it Physics of nearby Galaxies. Nature or Nurture?},
ed. T. Thuan, C. Balkowsky et J. Van, (Fronti\`eres: Gif sur Yvette),
pag. 3

\ref Butcher H., Oemler A, 1985, ApJS 57, 665

\ref Capaccioli, M., Vietri, M., Held, E.V., Lorentz, H., 1991, ApJ 371, 535

\ref Couch W., Ellis R., Sharples R. et al., 1994, ApJ 430, 121

\ref Doi M., Fukugita M., Okamura S. 1993, MNRAS 264, 832

\ref Dressler A., 1980, ApJS 42, 565

\ref Dressler A., Oemler A., Butcher H., Gunn J., 1994a, ApJ 430, 107

\ref Dressler A., Oemler A., Sparks W., Lucas R., 1994b, ApJ 435, L23

\ref Edge A., Steward A., 1991, MNRAS 252, 428

\ref Gavazzi G., Trinchieri G., 1989, ApJ 342, 718

\ref Hubble E., 1936, {\it The Real of the Nebulae},  New Haven: Yale 
University Press

\ref Kennicutt R., 1981, AJ 86, 1846

\ref Kormendy J., 1979, ApJ 227, 714

\ref Lahav O., Naim A., Buta R. et al., 1995, Science 267, 859

\ref Michard R., Marchal J. 1993, A\&AS 98, 29

\ref Michard R., 1996, Astroph. Lett. \& Comm., 31, 187

\ref Naim A., Lahav O., Buta R. et al., 1995, MNRAS, 274, 1107

\ref Nieto J.-L., Bender R., Poulain P., Surma P., 1992, 
A\&A 257, 97

\ref Nieto J.-L., Poulain P., Davoust E., 1994, A\&A 283, 1

\ref Poulain P., Nieto J.-L., Davoust E., 1992, A\&AS 95, 129

\ref Roberts M., Haynes M., 1994, ARA\&A 32, 115

\ref Rood H., Baum W., 1967, AJ 72, 398

\ref Saglia R., Bender R., Dressler A., 1993, A\&A 279, 77

\ref Sandage A., 1961,  {\it The Hubble Atlas of Galaxies}, 
(Washinton: Carnegie Institution)

\ref Sanrom\`a M., Salvador-Sol\'e E., 1990, ApJ 360, 16

\ref Sarazin C., 1986, {\it X-ray emissions from clusters of galaxies}, 
Cambridge University Press

\ref Schombert J., 1986, ApJS 60, 603

\ref Stanford S., Eisenhardt P., Dickinson M., 1995, ApJ 450, 512

\ref Staveley-Smith L, Davies R., 1988, MNRAS 231, 833

\ref Tammann G., 1987, IAU 124, 151

\ref van den Bergh S., 1990, Proc. ASP, Evolution of the 
Universe of Galaxies, Edwin Hubble Centennial Symposium, Ed. R.G. Kron, 
10, 70 

\ref de Vaucouleurs G., Buta R., 1980, ApJS 44, 451 

\ref de Vaucouleurs G., de Vaucouleurs A., Corwin H. et al., 1991 {\it 
Thirth Reference Catalogue of Bright Galaxies}
(New York: Springer Verlag), (RC3)

\ref Whitmore B., Gilmore D., 1991, ApJ 367, 64

\ref Whitmore B., Gilmore D., Jones C., 1993, ApJ 407, 489

\bye